\documentstyle[aps,prl,graphicx,amsbsy]{revtex}
\begin{document}
\draft
\hyphenation{scat-tering}
\wideabs{

\title{Formation of a magnetic soliton lattice in copper metaborate}

\author{B. Roessli$^1$, J. Schefer$^1$,  G. Petrakovskii$^2$, B. Ouladdiaf$^3$, M. Boehm$^{1,3}$, 
U. Staub$^4$, A. Vorontinov$^2$, and L. Bezmaternikh$^2$}
\address
{
$^1$Laboratory for Neutron Scattering, ETH Zurich \& Paul Scherrer Institute, CH-5232 Villigen PSI\\
$^2$Institute of Physics SB RAS, 660036 Krasnoyarsk, Russia\\
$^3$Institut Laue-Langevin, Av. des Martyrs, 38042 Grenoble, Cedex 9\\
$^4$Swiss Light Source, Paul Scherrer Institute,CH-5232 Villigen PSI, Switzerland
}


\maketitle

\begin{abstract}
\indent    
The magnetic ground-state of copper metaborate is 
investigated by means of elastic neutron scattering.
The magnetic structure of $\rm CuB_2O_4$ is incommensurate with respect to the 
chemical lattice at T=1.8K and undergoes a continuous 
phase transition to a non-collinear commensurate antiferromagnetic state which is 
realized at T=10K. Close to the phase transition higher-order magnetic
satellites are observed.  
Coexistence of long-range and 
short-range magnetic order is observed in both magnetic phases. This suggests
 that the association of  
the Dzyialoshinskii-Moriya interaction, lattice symmetry and tetragonal anisotropy leads to the formation 
of a three-dimensional magnetic soliton lattice. 
\end{abstract}
}

\noindent
Incommensurate magnetic structures are characterized by
modulations of their spin arrangements over periods which are long compared
to the size of the chemical cell and not commensurate with the underlying 
lattice \cite{chatt}. The existence of such magnetic structures  
in compounds  with localized spin densities is either due to 
competitions between exchange interactions or relativistic effects like
spin-orbit coupling. While the former effects can be found accidentally, 
the latter mechanism depends on lattice symmetry. Relativistic interactions 
 were first  
considered by Dzyaloshinskii \cite{dzyaloshinskii} and given a microscopic description 
 by Moriya \cite{moriya}. The Dzyaloshinskii-Moriya (DM) interaction is usually written
as the cross-product of interacting spins $H_{DM}=\bf{D\cdot(S_1 \times S_2)}$
where \textbf{D} is the DM-vector. The direction of the DM-vector
is determined by the bond symmetry and its scalar by the strength of the
spin-orbit coupling \cite{moriya}. The antisymmetric
DM-interaction in antiferromagnets favors a canted arrangement of the magnetic moments
which results in a weak spontaneous magnetization.
Spiral structures arise from the presence  
of a term linear in the gradient of the magnetization
(Lifshitz invariant \cite{lifschitz}) in the thermo-dynamical potential.
Dzyaloshinskii has shown that a Lifshitz invariant  is naturally obtained from 
  relativistic interactions in  chemical structures belonging to the 
 space groups  
$D_2, D_{2d}, C_{3h}, D_3, D_{3h}, S_4, D_4, D_6, T, T_d, O$ (Schoenflies notation).
The presence of an additional crystal anisotropy distorts
the simple helicoidal spin arrangement and the angle evolution of the
magnetic moments is described by  the solution of the sine-Gordon equation 
for a soliton lattice \cite{dz2,izyumov}. It should be recognized that a soliton corresponds 
to localized or topological dislocations of a periodic structure which arise when  
non-linear forces are present\cite{bishop}. 
In a magnetically ordered crystal, a soliton is a deformation 
of a regular spiral resulting in a non-uniform rotation of the magnetic moments along the 
pitch of the helix. 
In that case, Dzyialoshinskii 
has shown that a second order phase transition 
from the incommensurate to a commensurate structure occurs 
either at a given  temperature $T^\star$ or 
equivalently in the presence of a large magnetic field. 
While the formation of an incommensurate chemical lattice 
has been often observed in condensed matter physics \cite{bak}, and more 
recently in the spin-Peierls compounds TTF-CuBDT \cite{kiryukhin} 
and $\rm CuGeO_3$ \cite{kiryukhin2},
the presence of solitons in a three-dimensional magnetic lattice is unusual.  
 To our knowledge, a field-induced incommensurate-commensurate 
phase transition has been observed only 
for the Dzyaloshinskii-Moriya helimagnet $\rm Ba_2CuGe_2O_7$ \cite {zheludev}.  
\\
The bulk magnetic properties of $\rm CuB_2O_4$ were reported
 in ref. \cite{petrakovskii}. The specific heat data show 
two magnetic phase transitions at $\rm T_N=21K$ and $\rm T^\star$=10K. 
From magnetization and susceptibility measurements it is concluded that 
copper metaborate is a weak ferromagnet for $\rm T^\star \le T \le T_N$
 \cite{petrakovskii}.
In this letter we report elastic neutron scattering experiments
in copper metaborate $\rm CuB_2O_4$. In the 
temperature range $\rm 10K \le T \le 21K$ the magnetic structure is 
commensurate.
Upon lowering the temperature below $\rm T^\star$=10K, a commensurate to incommensurate 
 magnetic phase transition of the copper sublattice occurs. 
In addition, strong diffuse scattering is observed 
at the magnetic Bragg positions 
indicating that within the time window  accessible by neutrons 
spin fluctuations  coexist with static moments.  
Together with the fact that  Lifshitz invariant is allowed by 
 the lattice symmetry, this  suggests that 
copper metaborate forms a magnetic soliton lattice at low temperature. \\
The neutron measurements were performed at the neutron spallation source 
SINQ of the Paul-Scherrer Institut and at the high-flux neutron source of  
the Institut Laue-Langevin.
The single crystal  used in the present experiment was prepared using $\rm
^{11}B$ at the Institute of Physics Krasnoyarsk. 
The chemical structure was determined in the paramagnetic phase at T=25K. 
Due to the relatively large volume and low mosaic of the single crystal, 
it was necessary to 
correct the observed neutron intensities 
for extinction. To allow for a reliable 
correction, the neutron measurements were carried out at two wavelengths
 $\rm \lambda =2.4 \AA$ and $\rm 1.2  \AA$, respectively. 
 The refined parameters of the chemical structure 
were found to be in agreement with the X-ray 
results of Martinez \textit{et al.} \cite{martinez}. 
 $\rm CuB_2O_4$ crystallizes in 
space group  $I\bar 4 2 d$ ($D_{2d}^{12}$) with lattice constants a=11.528 
$\rm \AA$, c=5.607 $\rm \AA$. The unit cell contains 12 formula units. 
The $\rm Cu^{2+}$ ions occupy two 
inequivalent positions  at Cu(A)=(site 4b, point symmetry $\rm S_4$, $\rm 0 0 \frac{1}{2}$) 
and Cu(B)=(site 8d, point symmetry $\rm C_2$, $\rm  x \frac{1}{4} \frac{1}{8}$, 
x=0.0815), respectively. 
Cu(A) is at the center of a square unit formed by four oxygen ions, 
while Cu(B) is surrounded by six oxygen ions located at the vertices 
of a distorted octahedron. From high-resolution neutron powder diffraction 
 we conclude that $\rm CuB_2O_4$ does not undergo 
any structural phase transition down to T=1.5K.
Systematic scans in reciprocal space allowed magnetic 
reflections to be found at commensurate Bragg positions for T=12K. 
At this temperature, the observation of forbidden  reflections 
like (1,1,0) or (0,0,2) shows that the magnetic structure is 
antiferromagnetic. The magnetic structure is accordingly 
described by the propagation vector $\vec k=0$, 
so that the magnetic and chemical cells coincide.
As the lattice symmetry operation $I$ is also a magnetic translation when  
$\vec k=0$, the 
relevant irreducible representations of the magnetic structure are those of the 
point group $ \bar 4 2 m$. This point group contains eight elements and has five 
irreducible representations. Four of them are one-dimensional ($\rm \Gamma_1, \Gamma_2, 
\Gamma_3$ and $\rm \Gamma_4$) and one, labelled  $\rm \Gamma_5$, is two-dimensional.
 The reduction of the induction representation 
gives $\rm \Gamma_{4b}=\Gamma_3+\Gamma_4+2\Gamma_5$ and 
$\rm \Gamma_{8d}=\Gamma_1+2\Gamma_2+\Gamma_3+2\Gamma_4+3\Gamma_5$, respectively.
The magnetic modes 
$\rm \Gamma_3$ and $\rm \Gamma_4$  of the $\rm 4b$ site correspond 
to a collinear ferromagnetic and 
antiferromagnetic ordering along the z-axis, respectively. The modes associated 
with  the  
$\rm \Gamma_5$ representation describe a non-collinear magnetic 
structure with the magnetic moments 
rotated by 90$^\circ$ from each other. Similar magnetic modes for 
site 8d can be deduced again with help of group theory. 
From  a least-square 
refinement of a diffraction set consisting  of 
25 pure magnetic peaks ($\rm R_F=4\%$), we find that the magnetic structure 
of $\rm CuB_2O_4$ can be described as a non-collinear arrangement of both the 
Cu(A) and Cu(B)-spins along the diagonals of the tetragonal plane.  
The Cu(A) magnetic moments possess a small component  
$\mu_z=0.25\mu_B$ parallel to  the c-axis which corresponds to an angle of 
$\rm 14^\circ$ out-of the tetragonal ab-plane. 
Symmetry analysis of the chemical structure of $\rm CuB_2O_4$  
indicates that the DM-interaction 
is allowed between the Cu(A) nearest-neighbour spins\cite{moriya}. 
The DM-vector is  perpendicular to 
the tetragonal ab-plane of the crystal and, accordingly, the DM-interaction 
favors the non-collinear 
spin arrangement which is actually observed. This indicates that the DM-interaction 
plays a significant role in forming the magnetic ground-state in $\rm CuB_2O_4$.
From a simple scaling 
with the data taken in the paramagnetic phase, 
we obtain the value of the magnetic moment  
$\rm \sim1\mu_B$ 
for the Cu(A) spins at T=12K.  Within the precision of 
the present measurements, the Cu(B) spins are   
confined within the ab-plane and have a small magnetic moment 
 $\rm \mu\sim0.25\mu_B$, as shown in Fig.\ref{Fig1}. 
 As Cu(A) and Cu(B) magnetic moments do not compensate for each other, 
a spontaneous 
ferromagnetic moment equal to $0.1\mu_B$ per formula unit exists at T=12K.
\\
As shown in Fig.\ref{Fig2},  
the propagation vector $\vec k$ is temperature-dependent below T=10K and 
two magnetic satellites appear  
at symmetrical positions with respect to the commensurate reciprocal 
lattice-points.
This shows that the magnetic structure of $\rm CuB_2O_4$ 
 becomes incommensurate along the tetragonal 
axis. The period of the spin modulation continuously 
increases  from $\vec k=0$ below $\rm T^\star$=10K
 to  $\vec k=(0,0,0.15)$ at T=1.8K. At this temperature,
the modulation of the spin structure 
has a period of $\rm  c/0.15\sim 40 \AA $ 
along the crystallographic c-axis.   
The evolution of the satellite position with temperature is found to follow a 
power law
\begin{equation}
|\vec k(T)|\propto{{(T-T^\star)^\nu}}
\label{expo}
\end{equation}
with $\nu$=0.48. As shown in Fig.\ref{Fig2}, Eq.\ref{expo} describes the temperature dependence 
of the propagation vector in the incommensurate phase.  
As $\vec k(T)$  
 smoothly goes to zero with increasing temperature, the period of the helix growths 
to infinity at $\rm T^\star$=10K. 
In addition, considerable diffuse scattering 
superimposes on the resolution-limited Bragg
peaks for neutron scattering vectors \textbf{Q} along the [0,0,1]
crystallographic direction, as shown in Fig.\ref{Fig3}. The intensity of 
the diffuse scattering increases with increasing temperature and 
exhibits a critical divergence close to $T^\star$ (Fig.\ref{Fig4}), 
in accordance with  the observation of susceptibility and specific
heat peaks at this temperature \cite{petrakovskii}. 
The line-shape
of the diffuse scattering is well reproduced by 
the Fourier transform of the 
spin correlation function  
proposed by Ornstein and Zernicke \cite{collins}. This function, which is a Lorentzian 
\begin{equation}
S(\vec Q)={A\kappa \over {\kappa^2+Q^2}},
\label{difscat}
\end{equation}
is valid only in the temperature range 
where magnetic fluctuations are large. 
The average correlation length $\xi$ obtained from the
line-width  
$\xi= {1\over \kappa}$ yields $\rm \xi \sim 32 \AA$ at $T^\star$.  
$\kappa$  decreases upon passing into the 
incommensurate magnetic phase and reflects that 
the correlation length  
increases with decreasing temperatures up to $\rm \xi \sim 70\AA$ at T=1.8K. 
Diffuse scattering is unexpectedly 
observed at the lowest temperature reached in 
this experiment. Namely, for   
three-dimensionally ordered  magnets, critical fluctuations are usually  
sizable only in a small temperature range below 
the magnetic phase-transition. 
On the contrary, for spiral structures caused by relativistic 
interactions, a continuous intensity distribution is expected in a large
temperature range, due to perturbations in the simple helix
structure.  That diffuse scattering is observed at the lowest
temperature reached in this experiment and diverges when
$T\rightarrow T^\star$ demonstrates the validity of the
soliton concept for $\rm CuB_2O_4$. In that context, the soliton width is
given by the value of the parameter $\rm \xi$\cite{izyumov2}.
\\
Taking into account the crystalline anisotropy, the energy density 
 in the presence of an external magnetic field can be written as
\begin{eqnarray}
 F & = & J\bf{M}_1\bf{M}_2-D(\bf{M}_1 \times \bf{M}_2)_{z}  \nonumber  \\
  &   & - K_2(cos^2\theta_1+cos^2\theta_2)-K_4(cos^4\theta_1+cos^4\theta_2) \nonumber \\
  &   &
-K_4^{perp.}(cos^4\phi_1+cos^4\phi_2)-\bf{H}(\bf{M}_1+\bf{M}_2),
\label{energy}
\end{eqnarray}
 where J is the inter-sublattice exchange 
interaction; $\bf{M}_1$ and $\bf{M}_2$ are the magnetization vectors of the two sublattices; $\rm K_2$ and 
$\rm K_4$ correspond to the  second and fourth order uniaxial anisotropy constants; $\rm K_4^{perp.}$ 
represents the anisotropy in the tetragonal plane; 
$\theta$ and $\phi$ are the polar and azimuthal 
angles of the sublattice magnetization, respectively. 
Exchange interactions and anisotropy values in copper metaborate 
are known from magnetization and antiferromagnetic resonance experiments,  
$\rm 2M_0=160 G$, $\rm DM_0=1900 Oe$, $\rm H_E=JM_0= 2.9 \times 10^4 Oe$, $\rm K_2/M_0=21 Oe$, 
$\rm K_4/M_0=-8.4 Oe$, and  $\rm H_{perp.}=|K_4^{perp.}|/M_0 \sim 12 Oe$ at T=4.2K 
\cite{petrakovskii2,pankrats}.
The energy functional (\ref{energy}) describes the appearance of 
weak ferromagnetism in $\rm CuB_2O_4$ \cite{petrakovskii}. 
However, as $\vec k$  
decreases to zero when approaching $T^\star$ and diffuse scattering 
is observed  in a large  temperature range, and also as a phase transition is 
induced by applying a magnetic field 
in the tetragonal plane \cite{petrakovskii2}, 
competing exchange interactions do not appear to be the
main driving mechanism of 
the incommensurate-commensurate phase transition 
in copper metaborate. To explain this phase 
transition, it is reasonable to consider  
 the Lifshitz invariant in the energy density (\ref{energy})
\begin{equation}
F_l=\sigma(M_{1,x}\frac{dM_{2,y}}{dz}-M_{1,y}\frac{dM_{2,x}}{dz}),
\end{equation}
where $\sigma$ is the Lifshitz parameter.
 We note that the presence of the 
Lifshitz invariant  $F_l$ in the energy density of  $\rm CuB_2O_4$ 
is ensured 
both by the presence of the DM-interactions and the $D_{2d}$ symmetry of the lattice. 
The Lifshitz invariant $ F_l$  favors 
a magnetic ground-state 
given by the solution of the sine-Gordon equation. In the simplest case and in the absence of 
an external magnetic field, it   
 corresponds to a simple helix. 
 The helix structure can be distorted by the crystalline 
anisotropy which induces inhomogeneities in the magnetization density.
  The solution of the sine-Gordon equation 
results then in a magnetic soliton lattice \cite{izyumov2}. In that case, higher order 
satellites appear close the principal magnetic satellites. 
The amplitudes of these higher-order magnetic satellites increase with
temperature and are largest close to the 
incommensurate-commensurate phase transition \cite{izyumov3}. Fig.\ref{Fig5} 
shows a neutron scan along the (1,1,Q) direction at T=9K which reveals that 
when the temperature approaches $T^\star$, higher order satellites are
produced. We point out that the overall shape of the neutron scattering data  
differs from the theoretical calculations based on the classical Ginzburg-Landau functional
\cite{izyumov} for a simple lattice. 
This is probably due to the complexity of the chemical structure of copper 
metaborate. 
In any case, the presence of a multi-peak structure in the neutron 
spectrum is an indication 
that close to $\rm T^{\star}$, the magnetic structure of copper-metaborate can be viewed as 
a slightly distorted commensurate structure with domain walls.    
   
  As $\rm H_E \gg H_{perp.}$, it is 
possible to estimate the 
strength of the Lifshitz parameter from the expression 
for the wave vector of the incommensurate structure \cite{iz2}, namely 
$\rm \sigma M_0=k H_E\sim 4.3\times 10^3 Oe$ at T=1.8K. The temperature dependence of the propagation 
vector $\rm |\vec k(T)|$ and hence of the magnetic soliton length can be understood 
on qualitative grounds by noting that the 
anisotropy parameter of order $N$ in Eq.(\ref{energy}) depends on the magnetization as 
${K_N(T)\over K_N(0)} = [{ m(T)\over m(0)}]^{N(N+1)\over 2}$ \cite{rado}. 
As N is equal to 4 for tetragonal symmetry,   
a change in the magnetization curve 
by $\sim$10\% results in 
a decrease  of the anisotropy strength by a factor of $\sim$3 and hence in a corresponding increase 
of the helix vector.\\
In conclusion, we have shown that the magnetic structure of $\rm CuB_2O_4$ 
is incommensurate at low temperature. The 
propagation vector of the magnetic structure smoothly decreases to zero with increasing 
temperature. A non-collinear magnetic state commensurate with 
the underlying chemical lattice is eventually formed at 
$\rm T^\star$=10K.  Close to $T^\star$ higher-order magnetic satellites  and coexistence of 
short-range and long-range order is observed. This 
indicates that copper metaborate forms a three-dimensional magnetic soliton lattice. 
The origin of the incommensurate magnetic ground-state in $\rm CuB_2O_4$ is probably 
caused by the presence of a Lifshitz 
invariant in the thermo-dynamical potential. 

\begin{figure}
\centering
\vskip 4pt
\caption{Antiferromagnetic structure of $\rm CuB_2O_4$ in the 
commensurate phase.  The Cu(A) and Cu(B) positions are represented 
by black and open 
symbols respectively,  while the lengths and the 
directions of the arrows mirror  
 the $\rm Cu^{2+}$ magnetic moments $\rm \vec\mu$.}
\label{Fig1}
\end{figure}

\begin{figure}
\centering
\vskip 4pt
\caption{Temperature dependence of the propagation vector $\vec k$. The line
is the result of the calculation, as explained in the text. 
\textit{Inset}:  Neutron elastic scans along (3,3,Q) 
showing the evolution of 
the magnetic satellites in $\rm CuB_2O_4$ for selected temperatures. 
Below $T^\star$, the 
peak at the commensurate position is caused by multiple Bragg scattering.}
\label{Fig2}
\end{figure}

\begin{figure}
\centering
\vskip 4pt 
\caption{Elastic and diffuse scattering in $\rm CuB_2O_4$ below and above
the incommensurate-commensurate phase transition. Note the logarithmic
scale.The line is the result of a least-square fit to the data using a 
Gaussian function and Eq.\ref{difscat}.}
\label{Fig3}
\end{figure}

\begin{figure}
\centering
\vskip 4pt
\caption{Temperature dependence of the diffuse scattering in $\rm
CuB_2O_4$. Note the increase of intensity in the vicinity of
the incommensurate-commensurate phase transition. The line is to guide the
eye.}
\label{Fig4}
\end{figure}

\begin{figure}
\centering
\vskip 4pt
\caption{Neutron elastic scan along the (1,1,Q) direction at T=9 K showing
the presence of higher-order satellites in addition to the two principal magnetic
satellites. }
\label{Fig5}
\end{figure}

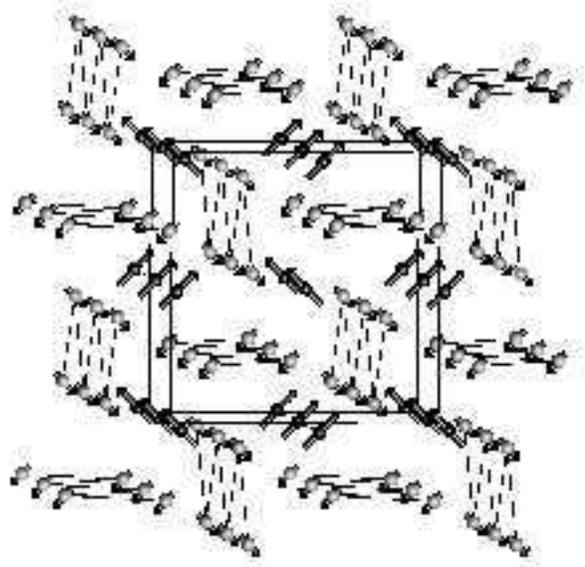
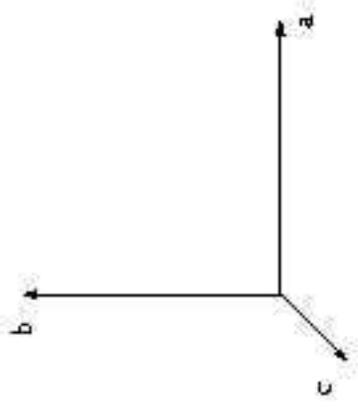
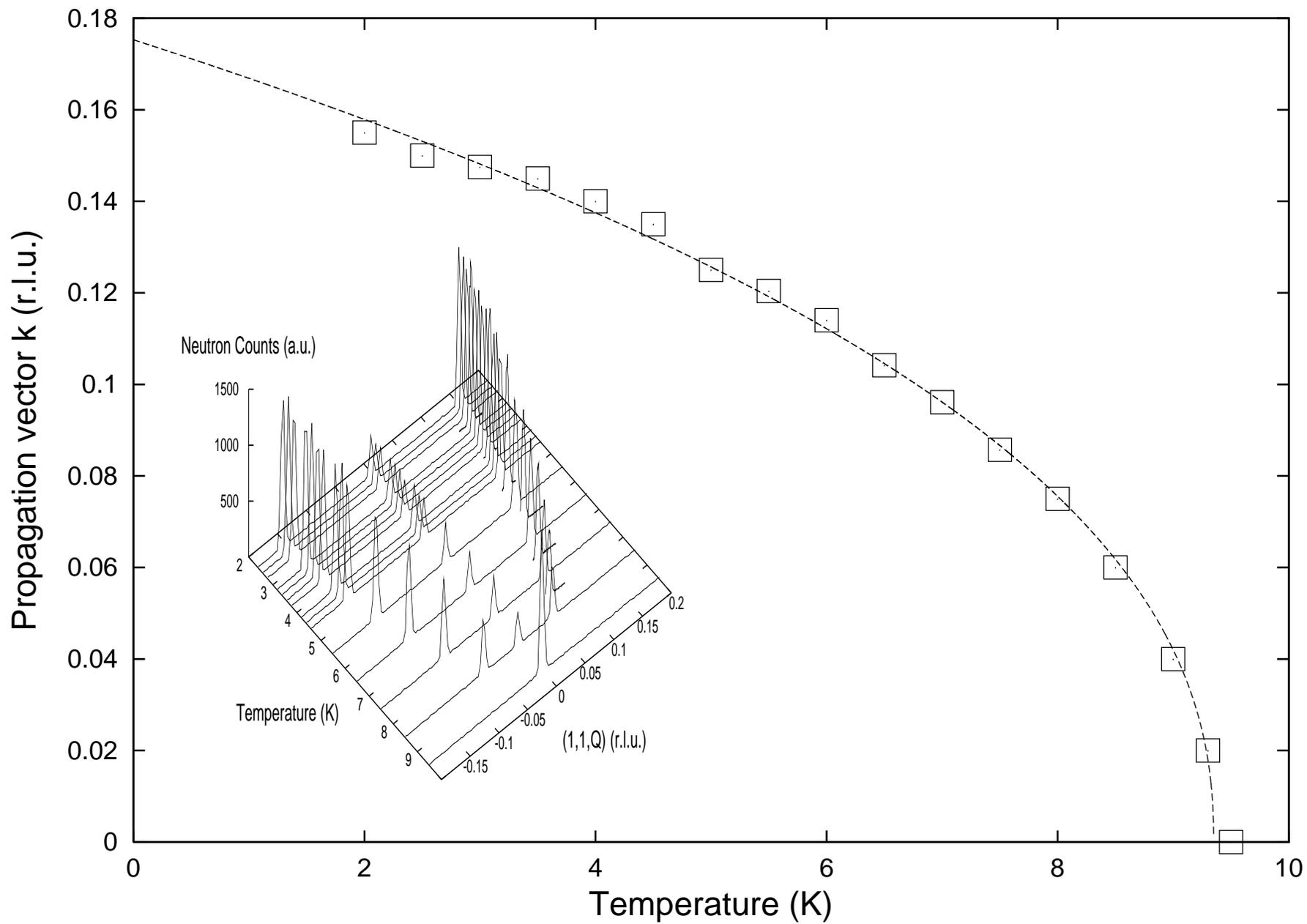
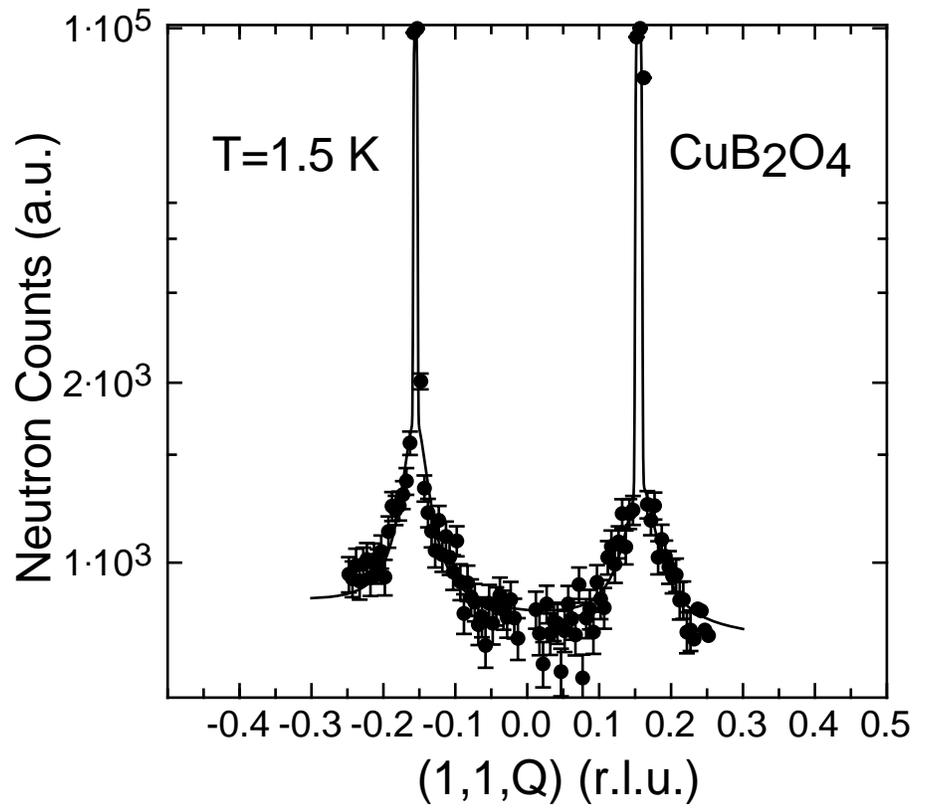
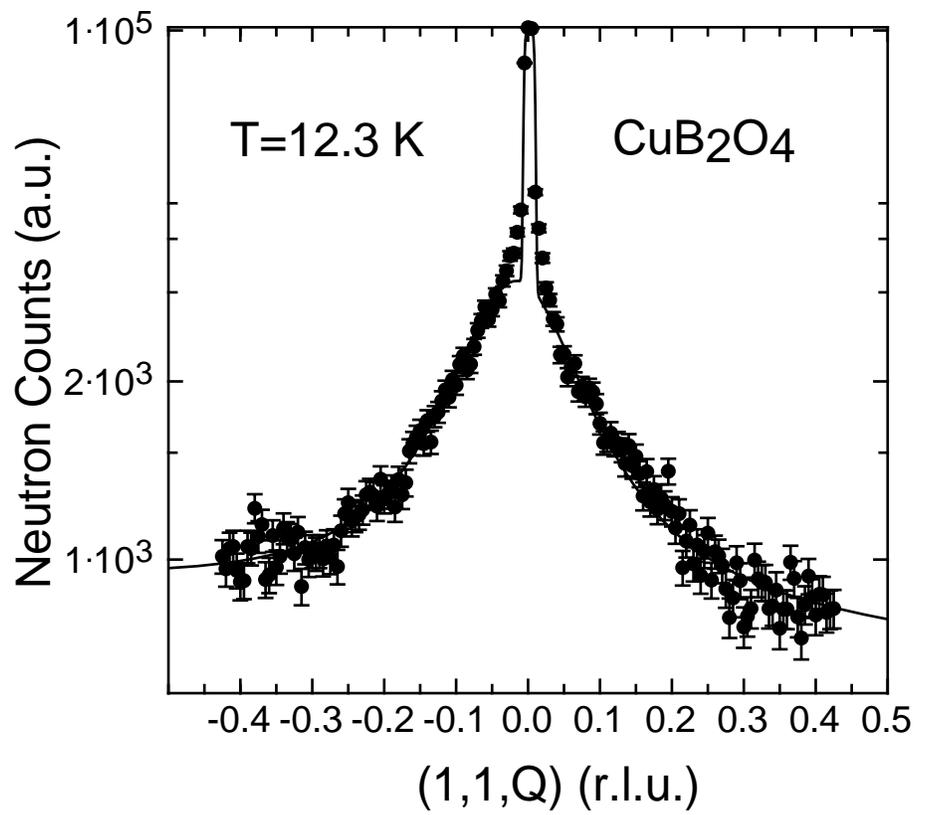
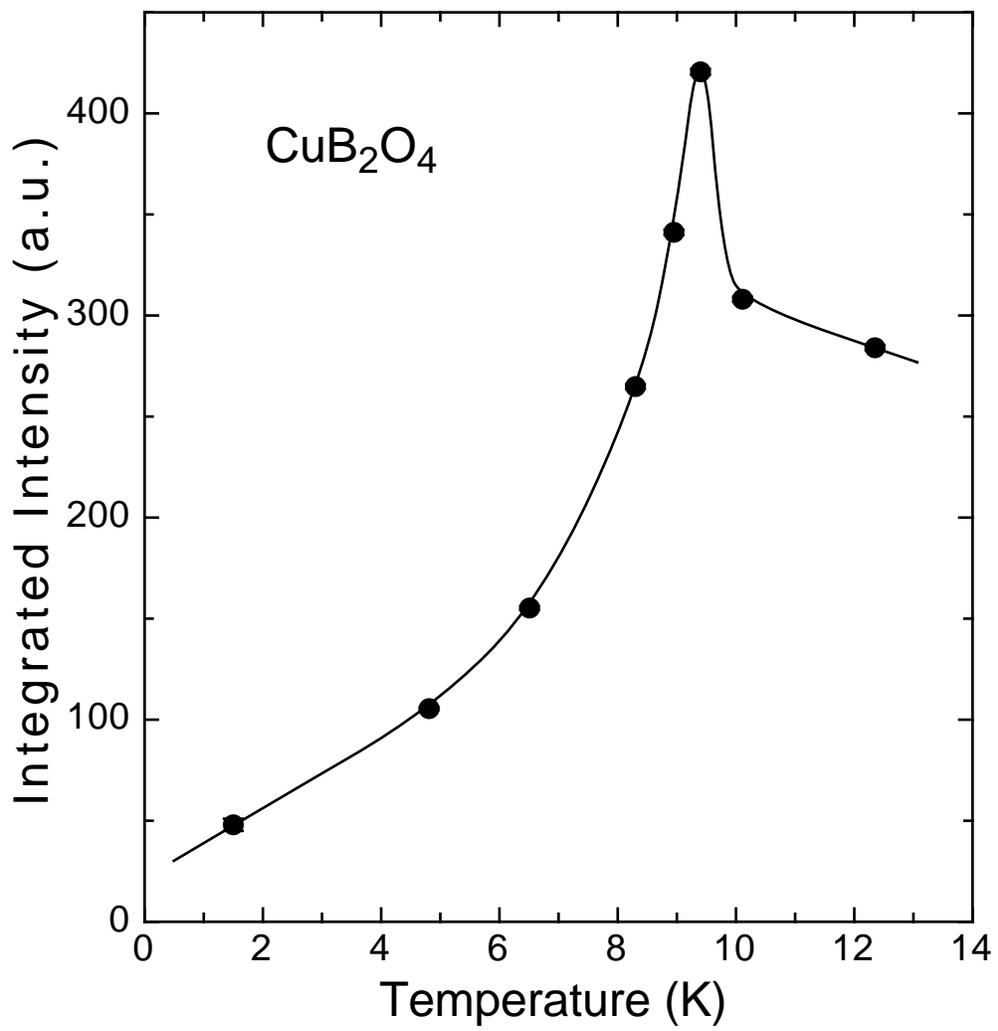
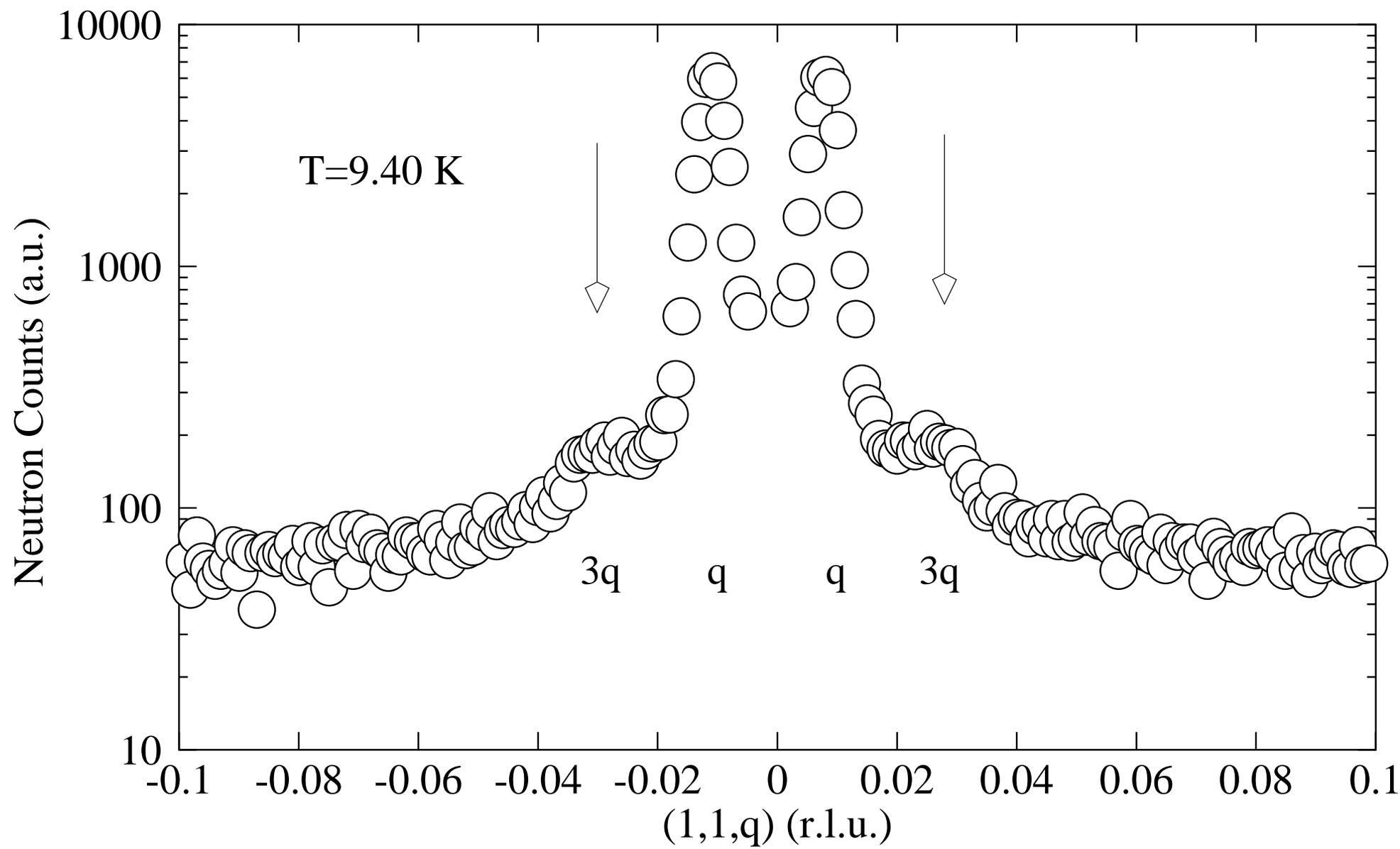
\end{document}